\documentclass[12pt]{article}
\usepackage{amsmath,amssymb}
\usepackage[body={18cm,20cm}]{geometry}
\newcommand{\R}{{\mathcal R}}

\newcommand{\da}{\delta_{1}}
\newcommand{\db}{\delta_{2}}
\newcommand{\dc}{\delta_{3}}
\newcommand{\dd}{\delta_{4}}
\renewcommand{\aa}{\alpha_{1}}
\newcommand{\ab}{\alpha_{2}}
\newcommand{\ac}{\alpha_{3}}
\newcommand{\ad}{\alpha_{4}}
\newcommand{\ba}{\beta_{1}}
\newcommand{\bb}{\beta_{2}}
\newcommand{\bc}{\beta_{3}}
\newcommand{\bd}{\beta_{4}}
\newcommand{\ga}{\gamma_{1}}
\newcommand{\gb}{\gamma_{2}}
\newcommand{\gc}{\gamma_{3}}
\newcommand{\gd}{\gamma_{4}}
\newcommand{\ma}{\mu_{1}}
\newcommand{\mb}{\mu_{2}}
\newcommand{\mc}{\mu_{3}}
\newcommand{\md}{\mu_{4}}
\newcommand{\me}{\mu_{5}}
\newcommand{\ra}{\rho_{1}}
\newcommand{\rb}{\rho_{2}}
\newcommand{\rc}{\rho_{3}}
\newcommand{\rd}{\rho_{4}}
\newcommand{\re}{\rho_{5}}
\newcommand{\na}{\nu_{1}}
\newcommand{\nb}{\nu_{2}}
\newcommand{\nc}{\nu_{3}}
\newcommand{\nd}{\nu_{4}}
\renewcommand{\ne}{\nu_{5}}
\newcommand{\sa}{\sigma_{1}}
\renewcommand{\sb}{\sigma_{2}}
\renewcommand{\sc}{\sigma_{3}}
\newcommand{\sd}{\sigma_{4}}

\def\Hs{{\mathsf H}}

\newcommand{\la}{\lambda} 
\newcounter{app}
\newcounter{sapp}[app]
\def\theapp{\Alph{app}}
\newcommand{\app}[1]{
\refstepcounter{app}{\vspace{7mm}
\noindent\Large\bf Appendix
\theapp.
 \ #1 \par \vspace{5mm}}
\setcounter{equation}{0}
\def\theequation{\Alph{app}.\arabic{equation}}}

\begin{document}
\title{Properties of the String Operator in the Eight-Vertex Model.}
\author{Klaus Fabricius\\Physics Department, University of Wuppertal\\
42097 Wuppertal, Germany.
\footnote{e-mail Fabricius@theorie.physik.uni-wuppertal.de}}
\maketitle
\begin{abstract}
\noindent
The construction of creation operators of exact strings in eigenvectors of the
eight vertex model at elliptic roots of unity of the crossing parameter which allow 
the generation of the complete set of degenerate eigenstates
is based on the conjecture that the 'naive' string operator vanishes. In this note
we present a proof of this conjecture. Furthermore we show that for 
chains of odd length the string operator is either proportional to the 
symmetry operator $S$ or vanishes depending on the precise
form of the crossing parameter.
 \end{abstract}
\maketitle
\section{Introduction}
The eight-vertex model of Baxter  is a lattice model whose transfer matrix is given by
\begin{equation}
{\bf T}_8(v)|_{\mu,\nu}={\rm Tr}W(\mu_1,\nu_1)
\label{transfer}
W(\mu_2,\nu_2)\cdots W(\mu_N,\nu_N)
\end{equation}
where $\mu_j,\nu_j=\pm1$ and $W(\mu,\nu)$ is a $2\times 2$ matrix whose
non vanishing elements are given as
\begin{eqnarray}
W(+1,+1)|_{+1,+1}&=W(-1,-1)|_{-1,-1}
=\rho\Theta(2\eta)\Theta(\lambda-\eta)\Hs(\lambda+\eta)=a(\lambda)
\nonumber\\
W(-1,-1)|_{+1,+1}&=W(+1,+1)|_{-1,-1}
=\rho\Theta(2\eta)\Hs(\lambda-\eta)\Theta(\lambda+\eta)=b(\lambda)
\nonumber\\
W(-1,+1)|_{+1,-1}&=W(+1,-1)|_{-1,+1}
=\rho \Hs(2\eta)\Theta(\lambda-\eta)\Theta(\lambda+\eta)=c(\lambda)
\nonumber\\
W(+1,-1)|_{+1,-1}&=W(-1,+1)|_{-1,+1}
=\rho \Hs(2\eta)\Hs(\lambda-\eta)\Hs(\lambda+\eta)=d(\lambda)
\label{bw8}
\end{eqnarray}
where $\Hs(u)$ and $\Theta(u)$ are Jacobi's Theta functions 
defined in  appendix \ref{Theta}. 
There are several paths leading to its solution. All have been either developed or
at least initiated by Baxter in a series of famous papers \cite{bax72}-\cite{bax733}.
The range of validity of the solutions depends on several parameters of the model:
It turns out to be essential whether the size $N$ of the lattice is even or odd.
Furthermore it is important whether the crossing parameter is generic or restricted to
elliptic 'root of unity' values. For details see \cite{bax72}-\cite{bax733} and 
\cite{fm1}-\cite{TQ1}. We only mention that the $TQ$ equation determines the eigenvalues 
of transfer matrix $T$ for even $N$ and unrestricted crossing parameter $\eta$ \cite{bax731}.
There exist well developed methods for the determination of eigenvectors of $T$ for 
even $N$ and 'root of unity' values of $\eta$
\begin{equation}
\eta = 2m_1K/L + im_2K'/L
\label{eta} 
\end{equation} 
See \cite{bax733},\cite{TakFadd79},\cite{Felder}.
Concerning the problem to obtain eigenvectors at generic $\eta$ 
information is given in \cite{Baxter2002}, footnote 18, in \cite{TakFadd79} after 
 (5.15) and in \cite{Felder} on page 497. 
In the following we restrict the crossing parameter $\eta$ to elliptic roots of unity
and for simplicity to the case $m_2=0$. 
Like in the six vertex model at root of unity
the transfer matrix of the eight-vertex model and the Hamiltonian of the related $XYZ$ 
spin chain have numerous degenerate multiplets of eigenvalues.
The symmetry algebra responsible for the degeneracies is well understood in the six vertex model
where it is the $sl_2$ loop algebra \cite{dfm}.
The problem to construct the operators which create the degenerate eigenvectors has been
solved in the six vertex model in \cite{Odyssey}. This solves simultaneously the problem 
to construct the current of the $sl_2$ loop symmetry studied in \cite{dfm}. The question arises if a similar
construction is necessary in the eight vertex model where the eigenvectors depend on free parameters
$s,t$ which have no influence on singlet states but affect the degenerate states.
For even $N$ eigenvectors of the transfer matrix are given by \cite{TakFadd79}
\begin{equation}
\psi_k =  \sum^{L-1}_{l=0}\exp(2\pi ilk/L)\prod_{m=1}^{n}B_{l+m,l-m}(\lambda_m,s,t)\Omega_{N}^{l-n}(s)
\label{psi1}
\end{equation}
where $\lambda_1,\cdots ,\lambda_n$ are Bethe roots.
Do eigenvectors obtained by variation of $s$ and $t$
span a complete degenerate subspace? That the answer is no has been shown in \cite{ellc}. To generate
the full degenerate multiplet a new string operator is needed:
the creation operator of a complete $B$-strings is for even $N$
\begin{equation}
B^{L_s,1}_{l}(\lambda_c)=\sum_{j=1}^{L_s}B_{l+1,l-1}(\lambda_1)\cdots
\left({\frac{\partial B_{l+j,l-j}}{ \partial \eta}}(\lambda_{j})
-\hat{Z}_{j}{\frac{\partial B_{l+j,l-j}} {\partial \lambda}}(\lambda_{j})\right)
\cdots B_{l+L_s,l-L_s}(\lambda_{L_s}) 
\label{Bstr}
\end{equation}
where the arguments $\lambda_k$ form an exact string of length $L_s$
\begin{equation}
\lambda_k = \lambda_c - 2(k-1)\eta, ~~~k=1,\cdots L_s
\label{lambdak}
\end{equation}
$\lambda_c$ is the string center.This is in addition to $s$ and $t$ a third free parameter.\\
${\hat Z}_1(\lambda_c)$ is defined in appendix \ref{ZXYP}.\\
This problem has also been studied in the framework of the Felder-Varchenko \cite{Felder} formalism by Deguchi \cite{deg3}.\\
We note that the string length $L_s$ and the integer $L$ occurring in (\ref{eta}) and (\ref{psi1}) are related by the rule \cite{ellc}
\begin{equation}
L_s = L {\rm ~~~~for~~ odd~~} L,\hspace{0.4 in} L_s = L/2 {\rm ~~~~for~~ even~~} L
\end{equation}
We now turn to the precise topic of this paper.
The construction of the string operator in \cite{ellc} rests on the conjecture
that the 'naive' string operator vanishes :
\begin{equation}
B_s = B_{l+1,l-1}(\lambda_1)\cdots B_{l+L_s,l-L_s}(\lambda_{L_s}) = 0
\label{BL}
\end{equation}
In  the six vertex model the equivalent relation is (see \cite{Tar})
\begin{equation}
B(\lambda_1)\cdots B(\lambda_{L_s}) = 0
\label{BL6}
\end{equation}
We intend to fill this gap by showing that (\ref{BL}) is satisfied for even $N$ in the eight vertex model.
Furthermore we find that for odd $N$ $B_s$ does NOT vanish but is given by a symmetry operator of the model. 
It is proportional to
\begin{equation}
S = \sigma^3\otimes \sigma^3\otimes \cdots \otimes \sigma^3
\label{S}
\end{equation}
provided that $2L_s\eta$ is an odd multiple of $2K$.If $\eta$ is an even multiple of $2K$
$B_s$ vanishes also for odd $N$.
\section{The formalism.}
In the following we work in the framework of the algebraic Bethe ansatz for the eight vertex model by Takhtadzhan and Faddeev \cite{TakFadd79}
and use their notation. For convenience some of their basic tools are listed in  appendix \ref{app1}.\\
We shall study the action of operators of type
\begin{equation}
{\cal{O}}^{N}_{j+1,l-1} = B^{N}_{j+1,l-1}(\lambda_1)\cdots B^{N}_{j+L_s,l-L_s}(\lambda_{L_s})
\label{Bstr1}
\end{equation}
on vectors denoted by $\Omega^{N}$.
The arguments $\lambda_k$ are defined in (\ref{lambdak}).
The superscript $N$ indicates the size of the one dimensional system. This is needed because our result
will be established recursively by relating systems of sizes $N$ and $N-1$.
$\Omega^N$ is any element of the set of $2^N$ independent basis vectors
\begin{equation}
\Omega^{N}=Z_{l_1}\otimes Z_{l_2}\otimes \cdots \otimes Z_{L_N}
\label{Zstate}
\end{equation}
where $Z_{l}$ stands for $X_{l}(\eta)$ or $Y_{l}(\eta)$.
$X_k(\lambda)$ and $Y_k(\lambda)$ are defined in  (\ref{Xk}) and (\ref{Yk}).
We note that in the algebraic Bethe-Ansatz \cite{TakFadd79} the eigenvectors of the transfer matrix are
obtained by the action of $B_{k,l}$ operators on the system of generating vectors defined in
\cite{TakFadd79}  (4.18) and (\ref{Omega}). The necessity to work in our case with more general
basis vectors arises from the intention to prove the operator relation (\ref{BL}).
This set of basis vectors is described in detail in (\ref{basis}).\\
In order to demonstrate the formalism to be used transparently we display the
string operator acting on a basis element of the space of states (in this example for a system of size $N=4$). What follows is nothing but a much more detailed presentation of the string operator (\ref{Bstr1}) acting on a basis vector
(\ref{Zstate}).
\begin{equation}
{\cal{O}}_{j+1,l-1} \Omega^{N} = 
\begin{array}{l l l l l l}
                      & Z_{l+2,\sa} & Z_{l+1,\sb}& Z_{l,\sc}& Z_{l-1,\sd}&     \\
m(\lambda_{L_s})^{-1}\tilde{Y}^{\ra}_{j+L_{s}}(\la_{L_{s}})&
 {\cal{L}}^{\sa,\rb}_{\da,\ra}  & {\cal{L}}^{\sb,\rc}_{\db,\rb}& {\cal{L}}^{\sc,\rd}_{\dc,\rc}& {\cal{L}}^{\sd,\re}_{\dd,\rd}&Y_{l-L_{s},\re}(\la_{L_s})\\
..........&..........&..........&..........&..........&..........\\
m(\lambda_{2})^{-1}\tilde{Y}^{\na}_{j+2}(\la_2)&
 {\cal{L}}^{\ga,\nb}_{\ba,\na}  & {\cal{L}}^{\gb,\nc}_{\bb,\nb}& {\cal{L}}^{\gc,\nd}_{\bc,\nc}& {\cal{L}}^{\gd,\ne}_{\bd,\nd}&Y_{l-2,\ne}(\la_2)\\
m(\lambda_{1})^{-1}\tilde{Y}^{\ma}_{j+1}(\la_1)&
 {\cal{L}}^{\ba,\mb}_{\aa,\ma}  & {\cal{L}}^{\bb,\mc}_{\ab,\mb}& {\cal{L}}^{\bc,\md}_{\ac,\mc}& {\cal{L}}^{\bd,\me}_{\ad,\md}&Y_{l-1,\me}(\la_1)\\
\end{array} 
\label{Llat}
\end{equation}
The ${\cal{L}}$ matrices appearing in the $k$ th row (counted from the bottom) depend on $\lambda_k$.
One easily recognizes the individual components of (\ref{Bstr1}) in the preceding expression.
A single row enclosed between $\tilde{Y}$ and $Y$ represents a $B$ operator written explicitly in terms of
local transition matrices as defined in
 (\ref{Bop}).
The leftmost operator in (\ref{Bstr1}) appears in the lowest row, the basis vector $\Omega^{N}$ in the top row.  
Expression (\ref{Llat}) can be either understood as row operators acting on the upper 
direct product of $Z_{l}$ vectors or as column operators which act on the direct product
of $Y_{l}$ vectors appearing in the rightmost column. We prefer the second choice as it 
takes from the outset into account that the spectral parameters $\lambda_k$ form a complete
string which leads to considerable simplifications. This is an essential part of our approach 
which enables us to relate different system sizes $N$ recursively.
The local transition matrix ${\cal{L}}$ is related to the R matrix by
\begin{equation}
{\cal{L}}^{\beta,\nu}_{\alpha,\mu}(\lambda) = {\cal{R}}^{\nu,\beta}_{\alpha,\mu}(\lambda,\eta)
\end{equation}
where $\alpha,\beta$ are quantum indices and $\mu,\nu$ auxiliary indices in the terminology 
introduced in \cite{TakFadd79}.
To apply the method of intertwining vectors we rewrite (\ref{Llat}) in terms of
the R matrix $\R$
\begin{equation}
{\cal{O}}_{j+1,l-1} = 
\begin{array}{l l l l l l}
                      & Z_{l+2,\sa} & Z_{l+1,\sb}& Z_{l,\sc}& Z_{l-1,\sd}&     \\
m(\lambda_{L_s})^{-1}\tilde{Y}^{\ra}_{j+L_{s}}(\la_{L_s})&
 \R^{\rb,\sa}_{\da,\ra}  & \R^{\rc,\sb}_{\db,\rb}& \R^{\rd,\sc}_{\dc,\rc}& \R^{\re,\sd}_{\dd,\rd}&Y_{l-{L_s},\re}(\la_{L_s})\\
..........&..........&..........&..........&..........&..........\\
m(\lambda_{2})^{-1}\tilde{Y}^{\na}_{j+2}(\la_2)&
 \R^{\nb,\ga}_{\ba,\na}  & \R^{\nc,\gb}_{\bb,\nb}& \R^{\nd,\gc}_{\bc,\nc}& \R^{\ne,\gd}_{\bd,\nd}&Y_{l-2,\ne}(\la_2)\\
m(\lambda_{1})^{-1}\tilde{Y}^{\ma}_{j+1}(\la_1)&
 \R^{\mb,\ba}_{\aa,\ma}  & \R^{\mc,\bb}_{\ab,\mb}& \R^{\md,\bc}_{\ac,\mc}& \R^{\me,\bd}_{\ad,\md}&Y_{l-1,\me}(\la_1)\\
\end{array} 
\label{BR}
\end{equation}
Note that the ${\cal{R}}$ matrices in the k-th row (counted from the bottom) depend on $\lambda_k$. 
For an application of a similar technique see \cite{DdVG}.

\section{Reduction of a column.}
We derive a relation between  ${\cal{O}}^{N}_{k,l}$ and  ${\cal{O}}^{N-1}_{k,l}$.
To accomplish this we remove the last column of (\ref{BR}) by making use of (\ref{ITV1})-(\ref{ITV12}).
Concentrating on this last column we find
\begin{equation}
\begin{array}{l}
X_{l-{L_s}+1,\sigma} \\
\R^{{\rho}',\sigma}_{\tau,\rho}Y_{l-{L_s},{\rho}'}(\la_{L_s})\\
\vdots\\
\R^{{\nu}_3',\epsilon}_{\delta,{\nu}_3}Y_{l-k-1,{\nu}_3'}(\la_{k+1})\\
\R^{{\nu}_2',\delta}_{\gamma,{\nu}_2}Y_{l-k,{\nu}_2'}(\la_{k})\\
\R^{{\nu}_1',\gamma}_{{\beta}_1,{\nu}_1}Y_{l-k+1,{\nu}_1'}(\la_{k-1})\\
\vdots\\
 \R^{\mu',\beta}_{\alpha,\mu}Y_{l-1,\mu'}(\la_1)\\
\end{array}
=\sum^{L_s}_{k=1}f_{k}
\left(\begin{array}{l}
Y_{l+1-{L_s},\rho}(\lambda_{L_s})\\
\vdots \\
Y_{l-k,\nc}(\lambda_{k+1})\\
X_{l+1-k,\nb}(\lambda_{k})\\
Y_{l-k,\na}(\lambda_{k-1})\\
\vdots\\
Y_{l-2,\mu}(\lambda_{1})\\
\end{array}\right)Y_{l-1,\alpha}(\eta)
+f_{0X}
\left(\begin{array}{l}
Y_{l+1-{L_s},\rho}(\lambda_{L_s})\\
\vdots \\
Y_{l-k,\nc}(\lambda_{k+1})\\
Y_{l-k+1,\nb}(\lambda_{k})\\
Y_{l-k+2,\na}(\lambda_{k-1})\\
\vdots\\
Y_{l,\mu}(\lambda_{1})\\
\end{array}\right)X_{l+1,\alpha}(\eta)
\label{col}
\end{equation}
This describes the result of the action of the last column of $\cal{R}$-operators on the right column of $Y$-vectors
in (\ref{BR}).
In this process the last column of $\cal{R}$-operators disappears (in accordance  with (\ref{ITV1})-(\ref{ITV8})) and 
the expression on the right hand side of (\ref{col}) replaces the right column of $Y$-vectors in (\ref{BR}). 
The $X$-vector on top
of the left hand side is converted to $Y_{l-1,\alpha}(\eta)$ and $X_{l+1,\alpha}(\eta)$ with free index $\alpha$. 
We note that on the left hand side $\alpha$ is the only free quantum index, the other indices 
from $\beta$ to $\sigma$ are
summed over whereas there are $L_s$ free auxiliary indices from $\mu$ to $\rho$. 
This simplifies on account of the intertwining 
relations (\ref{ITV1})-(\ref{ITV8}) to the direct products on the right side here written as 
columns in order to show clearly the origin
of each factor $Y$. 
The coefficients $f_{k},k=0,\cdots L_s$ follow after repeated use of (\ref{ITV1})-(\ref{ITV8}),
but we will obtain them more transparently from (\ref{col}).
To extract $f_m$ multiply (\ref{col}) by
\begin{equation}
\tilde{X}_{l}(\lambda_1)\otimes \tilde{X}_{l-1}(\lambda_2)\otimes \cdots \tilde{Y}_{l+1-m}(\lambda_m)\cdots \otimes 
\tilde{X}_{l+1-L-1}(\lambda_L)
\end{equation}
Note that all components are of type $\tilde{X}$ except $\tilde{Y}_{l-1+m}$.
Applying (\ref{XXYY}) and (\ref{XY}) we find
\begin{equation}
\begin{array}{lll}
&X_{l-{L_s}+1,\sigma}& \\
\tilde{X}^{\rho}_{l+1-{L_s}}(\lambda_{L_s})& \R^{{\rho}',\sigma}_{\tau,\rho}&Y_{l-{L_s},{\rho}'}(\la_{L_s})\\
\vdots\\
\tilde{X}^{\nc}_{l-m}(\lambda_{m+1})& \R^{{\nu}_3',\epsilon}_{\delta,{\nu}_3}&Y_{l-k-1,{\nu}_3'}(\la_{k+1})\\
\tilde{Y}^{\nb}_{l-m+1}(\lambda_{m})&\R^{{\nu}_2',\delta}_{\gamma,{\nu}_2}&Y_{l-k,{\nu}_2'}(\la_{k})\\
\tilde{X}^{\na}_{l-m+2}(\lambda_{m-1})&\R^{{\nu}_1',\gamma}_{{\beta}_1,{\nu}_1}&Y_{l-k+1,{\nu}_1'}(\la_{k-1})\\
\vdots\\
\tilde{X}^{\mu}_{l}(\lambda_1)&  \R^{\mu',\beta}_{\alpha,\mu}&Y_{l-1,\mu'}(\la_1)\\
\end{array}
\begin{array}{l}
= f_m \prod^{m-2}_{r=0}(\tilde{X}_{l-r}(\lambda_{r+1})Y_{l-r-2}(\lambda_{r+1}))\\
  ~~~~\times \prod^{L_s}_{s=m}m(\lambda_s) Y_{l-1,\alpha}(\eta)
\end{array}
\label{singlecol}
\end{equation}
The left hand side follows after repeated application of  (\ref{deltaX})-(\ref{deltaY}).
We find 
\begin{equation}
\begin{array}{lll}
&X_{l-{L_s}+1,\sigma}&\\
\tilde{X}^{\rho}_{l+1-{L_s}}(\lambda_{L_s})& \R^{{\rho}',\sigma}_{\tau,\rho}&Y_{l-{L_s},{\rho}'}(\la_{L_s})\\
\vdots\\
\tilde{X}^{\nc}_{l-m}(\lambda_{m+1})& \R^{{\nu}_3',\epsilon}_{\delta,{\nu}_3}&Y_{l-k-1,{\nu}_3'}(\la_{k+1})\\
\tilde{Y}^{\nb}_{l-m+1}(\lambda_{m})&\R^{{\nu}_2',\delta}_{\gamma,{\nu}_2}&Y_{l-k,{\nu}_2'}(\la_{k})\\
\tilde{X}^{\na}_{l-m+2}(\lambda_{m-1})&\R^{{\nu}_1',\gamma}_{{\beta}_1,{\nu}_1}&Y_{l-k+1,{\nu}_1'}(\la_{k-1})\\
\vdots\\
\tilde{X}^{\mu}_{l}(\lambda_1)&  \R^{\mu',\beta}_{\alpha,\mu}&Y_{l-1,\mu'}(\la_1)\\
\end{array}
\begin{array}{l}
=~~\prod^{m-1}_{j=1}h(\lambda_j+\eta)\prod^{{L_s}}_{j=m+1}h(\lambda_j-\eta)\\
~~\times \prod^{m-2}_{r=0}(\tilde{X}_{l-r}(\lambda_{r+1})Y_{l-r-2}(\lambda_{r+1}))\\
~~\times \prod^{L_s}_{s=1}m(\lambda_s)Y_{l-1,\alpha}(\eta)
\end{array}
\end{equation}
Result:
\begin{equation}
f_k(l) = h(2\eta)\frac{g(\tau_{l-k+1}+\lambda_k-\eta)}{g(\tau_{l-k+1})}\prod^{k-1}_{j=1}h(\lambda_j+\eta) \prod^{L_s}_{j=k+1}h(\lambda_j-\eta)
\label{fk}
\end{equation}
\begin{equation}
f_{0X} = \prod^{L_s}_{j=1}h(\lambda_j-\eta)
\label{f0X}
\end{equation}
To proceed further we insert (\ref{col}) into (\ref{BR}). 
\begin{eqnarray}
\label{BNBNM10}
&&B^{N}_{j+1,l-1}(\lambda_1)B^{N}_{j+2,l-2}(\lambda_2)\cdots B^{N}_{l+{L_s},l-{L_s}}(\lambda_{L_s})\Omega^N = \sum^{L_s}_{k=1}f_k\times \nonumber \\
&&B^{N-1}_{j+1,l-2}(\lambda_1)\cdots B^{N-1}_{j+k-1,l-k}(\lambda_{k-1})A^{N-1}_{j+k,l-k+1}(\lambda_k)
B^{N-1}_{j+k+1,l-k}(\lambda_{k+1})\cdots B^{N-1}_{j+{L_s},l+1-{L_s}}(\lambda_{L_s})\Omega^{N-1}\otimes Y_{l-1}(\eta) \nonumber \\
&&+f_{0X}B^{N-1}_{j+1,l}(\lambda_1)B^{N-1}_{j+2,l-1}(\lambda_2)\cdots B^{N-1}_{j+{L_s},l-{L_s}+1}(\lambda_{L_s})\Omega^{N-1}\otimes X_{l+1}(\eta) \nonumber \\
\end{eqnarray}
We commute the operator $A$ using relations (\ref{C2}) and (\ref{calpha}) until it is positioned directly in front of $\Omega$. 
\begin{eqnarray}
&&B^{N}_{j+1,l-1}(\lambda_1)B^{N}_{j+2,l-2}(\lambda_2)\cdots B^{N}_{l+{L_s},l-{L_s}}(\lambda_{L_s})\Omega^N = \nonumber \\
&&\left\{\sum^{L_s}_{k=1}(-1)^{{L_s}-k}f_k\prod^{{L_s}-1}_{m=k}\beta_{l-m}(\lambda_{m},\lambda_{m+1})\right\} \nonumber \\
&&B^{N-1}_{j+1,l-2}(\lambda_1)\cdots B^{N-1}_{j+{L_s}-1,l-{L_s}}(\lambda_{{L_s}-1})A^{N-1}_{j+{L_s},l-{L_s}+1}(\lambda_{L_s})\Omega^{N-1}\otimes Y_{l-1}(\eta) + \nonumber \\ 
&&f_{0X}B^{N-1}_{j+1,l}(\lambda_1)B^{N-1}_{j+2,l-1}(\lambda_2)\cdots B^{N-1}_{j+{L_s},l-{L_s}+1}(\lambda_{L_s})\Omega^{N-1}\otimes X_{l+1}(\eta) \nonumber \\
\label{BNBNM1}
\end{eqnarray}
The observation that the expression in braces vanishes is fundamental to that what follows. This will allow us to obtain our result
recursively.
So we prove that
\begin{equation}
F = \sum^{L_s}_{k=1}(-1)^{{L_s}-k}f_k(l)\prod^{{L_s}-1}_{m=k}\beta_{l-m}(\lambda_{m},\lambda_{m+1}) = 0.
\label{F0}
\end{equation}
We obtain from (\ref{calpha}) that
\begin{equation}
F = g(\tau_{l-{L_s}})\sum^{L_s}_{k=1}\frac{f_k(l)}{g(\tau_{l-k})}
\label{F1}
\end{equation}
and after insertion of (\ref{fk})
\begin{equation}
F = h(2\eta)g(\tau_{l-L_s})\sum^{L_s}_{k=1}\left\{\prod^{k-1}_{j=1}h(\lambda_j+\eta)\right\}\frac{g(\tau_{l-k+1}+\lambda_k-\eta)}{g(\tau_{l-k+1})g(\tau_{l-k})}
\left\{\prod^{L_s}_{j=k+1}h(\lambda_j-\eta)\right\}
\end{equation}
\begin{equation}
F\prod^{L_s-1}_{r=0}h(\tau_{l-r}) = h(2\eta)\sum^{L_s}_{k=1}\prod^{k-1}_{j=1}h(\lambda_j+\eta)\prod^{L_s}_{j=k+1}h(\lambda_j-\eta)
\prod^{k-2}_{r=0}h(\tau_{l-r})\prod^{L}_{r=k+1}h(\tau_{l-r})h(\tau_{l-k+1}+\lambda_k-\eta)
\end{equation}
We now insert (\ref{lambdak}) to make use of the fact that the arguments $\lambda_k$ form an exact string.
\begin{eqnarray}
&&F\prod^{L_s-1}_{r=0}h(\tau_{l-r}) = h(2\eta)\sum^{L_s}_{k=1}h(\tau_{l}+\lambda_c-(4k-3)\eta) \times \\ 
&&\left\{\prod^{k-1}_{j=1}h(\lambda_c-(2j-3)\eta))\right\}\left\{\prod^{L_s}_{j=k+1}h(\lambda_c-(2j-1)\eta)\right\}
\left\{\prod^{k-2}_{r=0}h(\tau_l-2r\eta))\right\}\left\{\prod^{L_s}_{r=k+1}h(\tau_{l}-2r\eta)\right\} \nonumber
\end{eqnarray}
We write this as
\begin{eqnarray}
F\prod^{L_s-1}_{r=0}h(\tau_{l-r}) = \sum^{L_s}_{k=1} p_k(\lambda_c)
\end{eqnarray}
A close inspection reveals that if for example $\lambda_{0}$ is  a zero of $p_1(\lambda_c)$ then $\lambda_{0}$ is also a zero of
$\sum^{L_s}_{k=2} p_k(\lambda_c)$. One finds that if $\lambda_0$ is a zero of $p_1$ all but two of the other $p_k$ have the same zero 
$\lambda_0$ and that the remaining two terms cancel for $\lambda_c=\lambda_0$.
Then $\sum^{L_s}_{k=2} \frac{p_k(\lambda_c)}{p_1(\lambda_c)}$ is doubly periodic and does not have poles.
It follows that it is a constant. It is easily shown that this constant is $-1$.
This proves that 
\begin{equation}
F = 0
\label{F}
\end{equation}
The corresponding column acting on a $Y$-type vector is
\begin{equation}
\begin{array}{l}
Y_{l-L_s-1,\sigma} \\
R^{{\rho}',\sigma}_{\tau,\rho}Y_{l-{L_s},{\rho}'}(\la_{L_s})\\
\vdots\\
\R^{{\nu}_3',\epsilon}_{\delta,{\nu}_3}Y_{l-k-1,{\nu}_3'}(\la_{k+1})\\
\R^{{\nu}_2',\delta}_{\gamma,{\nu}_2}Y_{l-k,{\nu}_2'}(\la_{k})\\
\R^{{\nu}_1',\gamma}_{{\beta}_1,{\nu}_1}Y_{l-k+1,{\nu}_1'}(\la_{k-1})\\
\vdots\\
\R^{\mu',\beta}_{\alpha,\mu}Y_{l-1,\mu'}(\la_1)\\
\end{array}
= f_{0Y}
\left(\begin{array}{l}
Y_{l-1-{L_s},\rho}(\lambda_{L_s})\\
\vdots \\
Y_{l-k-2,\nc}(\lambda_{k+1})\\
Y_{l-k-1,\nb}(\lambda_{k})\\
Y_{l-k,\na}(\lambda_{k-1})\\
\vdots\\
Y_{l-2,\mu}(\lambda_{1})\\
\end{array}\right)Y_{l-1,\alpha}(\eta)\\
\label{colY}
\end{equation}
\begin{equation}
f_{0Y} = \prod^{L_s}_{j=1}h(\lambda_j+\eta) 
\end{equation}
The explanatory remarks following (\ref{col}) apply also here.
We note that
\begin{equation}
\prod^{L_s}_{j=1}h(\lambda_j+\eta) = \pm\prod^{L_s}_{j=1}h(\lambda_j-\eta)
\label{prodh}
\end{equation}
where the minus sign holds if $2L_{s}\eta$ is an odd multiple of $2K$
and the plus sign if $2L_{s}\eta$ is a multiple of $4K$.
Because of (\ref{F0})  (\ref{BNBNM1}) is considerably simplified:
\begin{eqnarray}
\label{BX}
&&B^{N}_{j+1,l-1}(\lambda_1)B^{N}_{j+2,l-2}(\lambda_2)\cdots B^{N}_{j+{L_s},l-{L_s}}(\lambda_{L_s})\Omega^{N-1}\otimes X_{l+1-{L_s}}(\eta) =  \nonumber \\ 
&&f_{0X}B^{N-1}_{j+1,l}(\lambda_1)B^{N-1}_{j+2,l-1}(\lambda_2)\cdots B^{N-1}_{j+{L_s},l-{L_s}+1}(\lambda_{L_s})\Omega^{N-1}\otimes X_{l+1,\alpha}(\eta) \\
\nonumber
\end{eqnarray}
and from (\ref{colY}) follows directly
\begin{eqnarray}
\label{BY}
&&B^{N}_{j+1,l-1}(\lambda_1)B^{N}_{j+2,l-2}(\lambda_2)\cdots B^{N}_{j+{L_s},l-{L_s}}(\lambda_ {L_s})\Omega^{N-1}\otimes Y_{l-1-{L_s}}(\eta) = \nonumber \\
&&f_{0Y}B^{N-1}_{j+1,l-2}(\lambda_1)B^{N-1}_{j+2,l-3}(\lambda_2)\cdots B^{N-1}_{j+{L_s},l-{L_s}-1}(\lambda_{L_s})\Omega^{N-1}\otimes Y_{l-1,\alpha}(\eta) \\
\nonumber
\end{eqnarray}
We have found that the string operator acting on a chain of length $N$ is related
to a string operator acting on a chain of length $N-1$.
This  happens because the first column on the right hand side of (\ref{col}) which would destroy 
this simple relationship does not contribute on account of (\ref{F0}).
\section{The recursion process.}
To continue we introduce a shorter notation. For a column headed by an $X$ vector we write
\begin{equation}
{\cal{C}}_X(l) = {\cal{R}}(\lambda_1)^{\nu_1 \beta_1}_{\alpha \mu_1}{\cal{R}}(\lambda_2)^{\nu_2 \beta_2}_{\beta_1 \mu_2}\cdots{\cal{R}}(\lambda_{L_s})^{\nu_l \beta_{L_s}}_{\beta_{{L_s}-1 \mu_{Ls}}}X_{l \beta_{L_s}}(\eta)
\end{equation}
and for a column headed by a $Y$ vector
\begin{equation}
{\cal{C}}_Y(l) = {\cal{R}}(\lambda_1)^{\nu_1 \beta_1}_{\alpha \mu_2}{\cal{R}}(\lambda_1)^{\nu_2 \beta_2}_{\beta_1 \mu_2}\cdots{\cal{R}}(\lambda_{L_s})^{\nu_l \beta_{L_s}}_{\beta_{{L_s}-1 \mu_{L_s}}}Y_{l \beta_{L_s}}(\eta)
\end{equation}
The direct products of $Y$ and $X$ as occurring in (\ref{col}) will  be abbreviated by  
\begin{equation}
V_Y(l) = Y_{l}(\lambda_1)\otimes Y_{l-1}(\lambda_2)\otimes \cdots \otimes Y_{l-{L_s}+1}(\lambda_{L_s})
\end{equation}
\begin{equation}
V_{Y1X}(l,k) = Y_{l}(\lambda_1)\otimes Y_{l-1}(\lambda_2)\otimes \cdots \otimes Y_{l+2-k}(\lambda_{k-1})\otimes X_{l+3-k}(\lambda_{k})
\otimes Y_{l+2-k}(\lambda_{k+1})\otimes \cdots \otimes Y_{l+3-{L_s}}(\lambda_{L_s})
\end{equation}
where $k$ marks the position of $X$.
In this notation  (\ref{col}) and (\ref{colY}) read
\begin{equation}
{\cal{C}}_X(l-L_s+1)V_Y(l-1) = \sum^{L_s}_{k=1}f_kV_{Y1X}(l-2,k) \otimes Y_{l-1}(\eta) +f_{0X}V_{Y}(l)\otimes X_{l+1}(\eta)
\label{OXY}
\end{equation}
\begin{equation}
{\cal{C}}_Y(l-L_s-1)V_Y(l-1) = f_{0Y}V_{Y}(l-2)\otimes Y_{l-1}(\eta)
\label{OYY}
\end{equation}
We use the results obtained in the last section to study the action of B-strings on the set of vectors
described in (\ref{Zstate}). These states have the local structure
\begin{eqnarray}
\label{basis}
&&\cdots \otimes X_{l+1}(\eta) \otimes X_{l}(\eta) \otimes \cdots \\
&&\cdots \otimes Y_{l-1}(\eta) \otimes X_{l}(\eta) \otimes \cdots \nonumber \\
&&\cdots \otimes X_{l+1}(\eta) \otimes Y_{l}(\eta) \otimes \cdots \nonumber \\
&&\cdots \otimes Y_{l-1}(\eta) \otimes Y_{l}(\eta) \otimes \cdots \nonumber
\end{eqnarray}
i.e. the indices of neighboring vectors differ by $\pm 1$ depending on their order as shown in (\ref{basis}). 
This rule for the indices of consecutive $X$ and $Y$ guarantees that relation (\ref{ITV2}) is in all cases applicable.
In a system of length $N$ these vectors span a linear space of dimension $2^N$. They are obviously closely related to the family of vectors 
introduced by Baxter in \cite{bax733}  (1.3).
Then 
\begin{eqnarray}
\label{BN1}
&&\left(\prod^{L_s}_{k=1}m(\lambda_k)\right)B^{N}_{l+1,l-1}(\lambda_1)B^{N}_{l+2,l-2}(\lambda_2)\cdots B^{N}_{l+{L_s},l-{L_s}}(\lambda_{L_s})\Omega^{N}=  \\
&&V_{\tilde{Y}}(l+1){\cal{C}}_{Z_1}(l_1)\cdots {\cal{C}}_{Z_{N-1}}(l_{N-1}){\cal{C}}_X(l-L_s+1)V_Y(l-1) \nonumber
\end{eqnarray}
if $\Omega^{N}=\Omega^{N-1}\otimes X_{l+1-{L_s}}(\eta)$
and 
\begin{eqnarray}
\label{BN2}
&&\left(\prod^{L_s}_{k=1}m(\lambda_k)\right)B^{N}_{l+1,l-1}(\lambda_1)B^{N}_{l+2,l-2}(\lambda_2)\cdots B^{N}_{l+{L_s},l-{L_s}}(\lambda_{L_s})\Omega^{N}= \\
&&V_{\tilde{Y}}(l+1){\cal{C}}_{Z_1}(l_1)\cdots {\cal{C}}_{Z_{N-1}}(l_{N-1}){\cal{C}}_Y(l-L_s-1)V_Y(l-1) \nonumber
\end{eqnarray}
if $\Omega^{N}=\Omega^{N-1}\otimes Y_{l-1-{L_s}}(\eta)$ \\
We have shown (see  (\ref{BNBNM1}) and (\ref{F0})) that  (\ref{OXY}) and (\ref{OYY}) if inserted into B-strings effectively read
\begin{equation}
{\cal{C}}_X(l-L_s+1)V_Y(l-1) = f_{0X}V_{Y}(l)\otimes X_{l+1}(\eta)
\end{equation}
\begin{equation}
{\cal{C}}_Y(l-L_s-1)V_Y(l-1) = f_{0Y}V_{Y}(l-2)\otimes Y_{l-1}(\eta)
\end{equation}
where $f_{0Y} = \pm f_{0X}$
\subsection{The leftmost column.}
The last step in the recursive determination is the treatment of the first column.\\
{\bf Case a:}\\
If
\begin{equation}
\Omega^{N} = X_{l+n_x-n_y-{L_s}}\otimes \cdots \otimes X_{l+1-{L_s}} \hspace{0.3 in} {\rm ~~~or~~~}
\Omega^{N} = X_{l+n_x-n_y-{L_s}}\otimes \cdots \otimes Y_{l-1-{L_s}}
\label{OXX}
\end{equation}
the first column is
\begin{equation}
{\cal{C}}_X(l+n_x-n_y-L_s)V_Y(l+n_x-n_y-2)
\label{C1X}
\end{equation}
{\bf Case b:}\\
and if
\begin{equation}
\Omega^{N} = Y_{l+n_x-n_y-{L_s}}\otimes \cdots \otimes X_{l+1-{L_s}} \hspace{0.3 in} {\rm ~~~or~~~}
\Omega^{N} = Y_{l+n_x-n_y-{L_s}}\otimes \cdots \otimes Y_{l-1-{L_s}}
\label{OYX}
\end{equation}
the first column is
\begin{equation}
 {\cal{C}}_Y(l+n_x-n_y-L_s)V_Y(l+n_x-n_y)
\label{C1Y}
\end{equation}
In (\ref{OXX}) and (\ref{OYX}) the gap between the leading and  the trailing $X$ or $Y$ is filled with vectors having indices
according to the rule illustrated by (\ref{basis}). 
\subsubsection{Case a.}
From (\ref{OXY}) follows
\begin{eqnarray}
&&V_{\tilde{Y}}(l+1){\cal{C}}_X(l+n_x-n_y-L_s)V_Y(l+n_x-n_y-2) =   \\
&&\sum^{L_s}_{k=1}f_k(l+M-1)(V_{\tilde{Y}}(l+1)V_{Y1X}(l+M-3,k))\otimes Y_{l+M-2}(\eta) \nonumber \\
&&+f_{0X}(V_{\tilde{Y}}(l+1)V_Y(l+M-1))\otimes X_{l+M}(\eta) \nonumber
\label{ColX}
\end{eqnarray}
where $M = n_x-n_y$
\begin{eqnarray}
&&(V_{\tilde{Y}}(l+1)V_{Y1X}(l+M-3,k))= \\ 
&&\left\{\prod^{k-1}_{m=1}\tilde{Y}_{l+m}(\lambda_m)Y_{l+M-2-m}(\lambda_m)\right\}
\tilde{Y}_{l+k}(\lambda_k)X_{l+M-k}(\lambda_k)
\left\{\prod^{L_s}_{m=k+1}\tilde{Y}_{l+m}(\lambda_m)Y_{l+M-m}(\lambda_m)\right\} \nonumber
\label{ColM}
\end{eqnarray}
This expression is most conveniently handled once we recognize that it is
\begin{eqnarray}
&&(V_{\tilde{Y}}(l+1)V_{Y1X}(l+M-3,k))= \prod^{L_s}_{k=1}m(\lambda_k)\\ 
&&\left\{\prod^{k-1}_{m=1}B^{(0)}_{l+m,l+M-2-m}(\lambda_m)\right\}A^ {(0)}_{l+k,l+M-k}(\lambda_k)
\left\{\prod^{L_s}_{m=k+1}B^ {(0)}_{l+m,l+M-m}(\lambda_m)\right\}
\label{ColB0}
\end{eqnarray}
for a chain of length zero and that the permutation relations for $A$ and $B$ still hold in
this case (see (\ref{C20}) in Appendix \ref{Perm} ).
\begin{eqnarray}
&&(V_{\tilde{Y}}(l+1)V_{Y1X}(l+M-3,k))= \prod^{L_s}_{k=1}m(\lambda_k)\\
&&(-1)^{{L_s}-k}\prod^{L_s}_{k+1}\beta_{l+M-m}(\lambda_{m-1},\lambda_m)\prod^{{L_s}-1}_{m=1}B^{(0)}_{l+m,l+M-2-m}(\lambda_m)
A^{(0)}_{l+{L_s},l+M-{L_s}}(\lambda_{L_s})
\label{ColB01}
\end{eqnarray}
Insert  $\beta_{l+M-m}(\lambda_{m-1},\lambda_m)$ using (\ref{calpha}) 
\begin{eqnarray}
&&(V_{\tilde{Y}}(l+1)V_{Y1X}(l+M-3,k))= \prod^{L_s}_{k=1}m(\lambda_k)\\
&&\frac{g(\tau_{l+M-{L_s}-1})}{g(\tau_{l+M-k-1})}\prod^{{L_s}-1}_{m=1}B^{(0)}_{l+m,l+M-2-m}(\lambda_m)A^{(0)}_{l+{L_s},l+M-{L_s}}(\lambda_{L_s})
\label{ColB02}
\end{eqnarray}
It follows for the first term on the right hand side of (\ref{ColX}) 
\begin{eqnarray}
&&\sum^{L_s}_{k=1}f_k(l+M-1)(V_{\tilde{Y}}(l+1)V_{Y1X}(l+M-3,k))\otimes Y_{l+M-2}(\eta) = \prod^{L_s}_{k=1}m(\lambda_k)\\
&&\sum^{L_s}_{k=1}f_k(l+M-1)\frac{g(\tau_{l+M-{L_s}-1})}{g(\tau_{l+M-k-1})}
\prod^{{L_s}-1}_{m=1}B^{(0)}_{l+m,l+M-2-m}(\lambda_m)A^{(0)}_{l+{L_s},l+M-{L_s}}(\lambda_{L_s}) \otimes Y_{l+M-2}(\eta) \nonumber
\end{eqnarray}
We have already shown that (see (\ref{F1}))
\begin{equation}
\sum^{L_s}_{k=1}f_k(l+M-1)\frac{g(\tau_{l+M-{L_s}-1})}{g(\tau_{l+M-k-1})} = 0
\end{equation}
Therefore  (\ref{ColX}) is reduced to
\begin{equation}
V_{\tilde{Y}}(l+1){\cal{C}}_X(l+n_x-n_y-L_s)V_Y(l+n_x-n_y-2) = 
f_{0X}(V_{\tilde{Y}}(l+1)V_Y(l+M-1))\otimes X_{l+M}(\eta)
\label{ColX1}
\end{equation}
We claim that for even $N$
\begin{equation}
(V_{\tilde{Y}}(l+1)V_Y(l+M-1)) =\prod^{L_s}_{k=1}\tilde{Y}^{\mu}_{l+k}(\lambda_k)Y_{l+M-k,\mu}(\lambda_k) = 0
\end{equation}
The last expression vanishes if for some $1 \leq k \leq L_s$ 
\begin{equation}
2k\eta = (M-k)2\eta + 4rK
\label{k}
\end{equation}
or
\begin{equation}
k = M/2 + r_1L_s
\end{equation}
For even $N$ also $M$ is even and by appropriately adjusting $r_1$ 
an integer $k$ is found which satisfies (\ref{k}).
\subsubsection{Case b.}
From (\ref{OYY}) follows
\begin{eqnarray}
&&V_{\tilde{Y}}(l+1){\cal{C}}_Y(l+M-L)V_Y(l+M) =  \nonumber \\
&&f_{0Y}(V_{\tilde{Y}}(l+1)V_Y(l+M-1))\otimes Y_{l+M}(\eta)
\label{ColY}
\end{eqnarray}
In the same manner as above we can show that this is zero if $M$ is even.\\
This completes the proof of equation (\ref{BL}) for even $N$.
\begin{equation}
B_{l+1,l-1}(\lambda_1)\cdots B_{l+L_s,l-L_s}(\lambda_{L_s}) = 0
\label{BL1}
\end{equation}
\section{Chains of odd length.}
We remark that in the algebraic Bethe Ansatz eigenvectors are obtained for even $N$.
The eight vertex model for chains of odd length $N$ is nevertheless interesting as demonstrated in
\cite{fm3}, \cite{BazMang1}, \cite{BazMang2}. 
We therefore present also our result for odd $N$ which surprisingly shows that the string operator
does not vanish but is proportional to a simple symmetry operator.
In this case we have to collect the contributions of each column in the stepwise reduction
of a system of size $=N$ to a system of size $=1$ and to multiply finally by the result for the remaining
system of size $= 1$ obtained in the preceeding section.
We observe that the explicit form of $X$ and $Y$ given in (\ref{Xk}) and (\ref{Yk}) shows that for integer $m$
\begin{equation}
X_{l-{L_s}}(\eta) = -\sigma_3X_l(\eta) \hspace{0.4 in} Y_{l-{L_s}}(\eta) = \sigma_3Y_l(\eta) 
\hspace{0.4 in} {\rm if}~~~ 2L_s\eta=(2m+1)2K
\label{sigma3XY}
\end{equation}
\begin{equation}
X_{l-{L_s}}(\eta) = X_l(\eta) \hspace{0.65 in} Y_{l-{L_s}}(\eta) = Y_l(\eta) 
\hspace{0.58 in} {\rm if}~~~ 2L_s\eta=m4K
\label{II}
\end{equation}
\subsection{The case $2L_s\eta=(2m+1)2K$.}  
We conclude from  (\ref{BX}), (\ref{BY}),(\ref{ColX}),(\ref{ColY}) and (\ref{sigma3XY})
that the action of the string operator $B_s$ on a state $\Omega(n_X,n_Y)$ characterized in (\ref{Zstate}) and (\ref{basis})
where $n_X$ and $n_Y$ denote the number of $X$- and $Y$-vectors is given by 
\begin{equation}
B_{l+1,l-1}(\lambda_1)\cdots B_{l+L_s,l-L_s}(\lambda_{L_s})\Omega(n_X,n_Y) =
-\left(\prod^{L_s}_{k=1}m(\lambda_k)^{-1}\right)C(\lambda_c,M)f^{N}_{0X}\sigma_3\otimes \sigma_3 \otimes \cdots \otimes \sigma_3 \Omega(n_X,n_Y)
\label{Bodd1}
\end{equation}
where 
\begin{equation}
C(\lambda_c,M) = (V_{\tilde{Y}}(l+1)V_Y(l+M-1))=\prod^{L_s}_{k=1}\tilde{Y}^{\mu}_{l+k}(\lambda_k)Y_{l+M-k,\mu}(\lambda_k)
\label{C}
\end{equation}
and where (\ref{sigma3XY}) and $f_{0Y}=-f_{0X}$ as well as $n_X+n_Y = N$ and that N is odd have been taken into account.\\
\subsection{The case $2L_s\eta=m4K$}
We use (\ref{II}) and $f_{0Y}=+f_{0X}$ :
\begin{equation}
B_{l+1,l-1}(\lambda_1)\cdots B_{l+L_s,l-L_s}(\lambda_{L_s})\Omega(n_X,n_Y) =
\left(\prod^{L_s}_{k=1}m(\lambda_k)^{-1}\right)C(\lambda_c,M)f^{N}_{0X} \Omega(n_X,n_Y)
\label{Bodd2}
\end{equation}  
We claim that $C(\lambda_c,M)$ does not depend on $M=n_X-n_Y$. \\
Proof:
From (\ref{C}) and appendix \ref{app1} follows\\
\begin{equation}
C(\lambda,M) = \prod^{L_s}_{k=1}\frac{1}{g(\tau_{l+k})g(\tau_{l+M-k})}(\Theta(u)\Hs(v)-\Hs(u)\Theta(v))
\label{C1}
\end{equation}
with
\begin{equation}
u = t+2(l+k)\eta + \lambda_k \hspace{0.3 in} v = t + 2(l+M-k)\eta + \lambda_k \hspace{0.3 in} 
\lambda_k = \lambda_c - 2(k-1)\eta
\end{equation}
We rewrite this by using the relations of appendix \ref{Theta}
\begin{eqnarray}
&&C(\lambda,M) = \frac{2}{\Hs(K)\Theta(K)}\prod^{L_s}_{k=1}\frac{1}{g(\tau_{l+k})g(\tau_{l+M-k})}\Hs((M/2-k)2\eta)\Theta((M/2-k)2\eta) \times \nonumber\\ 
&&\Hs(t+\lambda_c+(l+1+M/2-k)2\eta +K)\Theta(t+\lambda_c+(l+1+M/2-k)2\eta +K) 
\label{C1a}
\end{eqnarray}
We compare $C(\lambda,M)$ with $C(\lambda,M+2)$. If $\eta = 2m_1K/L$
\begin{equation}
\prod^{L_s}_{k=1}g(\tau_{l+M+2-k}) = \prod^{{L_s}-2}_{k=-1}g(\tau_{l+M-k})= \prod^{L_s}_{k=1}g(\tau_{l+M-k})
\end{equation} 
as an even number of shifts by $2K$ occur.
Similarly a shift of $M$ by $2$ is compensated by a shift of $k$ by $1$ in
\begin{equation} 
\prod^{L_s}_{k=1}\Hs((M/2-k)2\eta)\Theta((M/2-k)2\eta)\Hs(t+\lambda_c+(l+1+M/2-k)2\eta +K)\Theta(t+\lambda_c+(l+1+M/2-k)2\eta +K)
\label{prodTh}
\end{equation}
It follows
\begin{equation}
C(\lambda_c,M+2) = C(\lambda_c,M)  
\end{equation}
and as $M$ is odd the simplest choice ist
\begin{equation}
C(\lambda_c,M) = C(\lambda_c,-1) =\prod^{L_s}_{k=1}\tilde{Y}^{\mu}_{l+k}(\lambda_k)Y_{l-1-k,\mu}(\lambda_k)
\end{equation}
Because of the presence of $\Hs((M/2-k)2\eta)$ in (\ref{prodTh})  $C(\lambda,M)$ vanishes if $2L_s\eta= m4K$ for integer m.
However it does {\em{not vanish}} if  $2L_s\eta= (2m+1)2K$
This means that for odd $N$ \\
$B_{l+1,l-1}(\lambda_1)\cdots B_{l+L_s,l-L_s}(\lambda_{L_s})$ is  proportional
to the operator $S=\sigma_3\otimes\sigma_3\cdots\otimes\sigma_3$.
\section{Summary of Results}
We have studied the properties of the operator 
\begin{equation}
B_s(\lambda_c) = B_{l+1,l-1}(\lambda_1)\cdots B_{l+L_s,l-L_s}(\lambda_{L_s})
\label{BLS}
\end{equation}
at roots of unity 
\begin{equation}
\eta = 2m_1K/L
\label{etax}
\end{equation}
1. For even $N$: $B_s=0$.
\\
2. For odd $N$ and $2L_s\eta \equiv~0 ~ ({\rm mod}~~4K)$:  $B_s=0$ \\ \\
3. For odd $N$ and $2L_s\eta \equiv~ 2K ~({\rm mod} ~~ 4K)$: 
\begin{equation}
B_{l+1,l-1}(\lambda_1)\cdots B_{l+L_s,l-L_s}(\lambda_{L_s})\Omega(n_X,n_Y) =
f(\lambda_c)\sigma_3\otimes \sigma_3 \otimes \cdots \otimes \sigma_3 \Omega(n_X,n_Y)
\label{Bodd3}
\end{equation}
\begin{equation}
f(\lambda_c) = -\left(\prod^{L_s}_{k=1}m(\lambda_k)^{-1}\right)f^{N}_{0X}\prod^{L_s}_{k=1}\tilde{Y}^{\mu}_{l+k}(\lambda_k)Y_{l-1-k,\mu}(\lambda_k)
\end{equation}
To clarify what happens for odd $N$ we give some examples.\\
If $\eta=K/3$ then because of (\ref{etax}) $L=6$ and $L_s=3$, $2L_s\eta=2K$ and (\ref{Bodd3}) applies.\\
If $\eta=2K/3$ then $L=3$ and $L_s=3$, $2L_s\eta=4K$ and $B_s=0$\\
If $\eta=K/2$ then $L=4$ and $L_s=2$, $2L_s\eta=2K$ and (\ref{Bodd3}) applies.\\
\section{Conclusions}
A common feature of the six vertex model and the eight vertex model for crossing parameters
at roots of unity is the occurrence of degneracies in the eigenvalues of the transfer 
matrix $T$ and the Hamiltonian $H$ .
This is by no means an exceptional case, because analytic solutions of the eight-vertex model which include the eigenvectors
of the transfer matrix $T$ exist only for  root of unity $\eta$.
For the six vertex model the underlying symmetry algebra is fairly well understood.
It has been shown in \cite{dfm} that the
$sl_2$ loop algebra symmetry is responsible for the degeneracies of eigenvalues
and its Chevalley generators have been explicitly constructed. 
The picture was completed by the construction of the generating
function of the operators in the mode basis (the current) in \cite{Odyssey} which allowed the determination
of the evaluation parameters.
This current plays two different roles: it generates the $sl_2$ loop algebra operators in the 
mode basis and it is the operator which creates exact Bethe strings in the algebraic Bethe ansatz.
It turns out that the set of B operators is itself not complete in the sense that it is not 
capable of creating the complete eigenstate of $T$. To achieve this the current operators have to
be incorporated.\\
We now turn to the eight vertex model. The result of this paper given in the summary completes
the construction of the operator which generates complete Bethe strings and is in this respect
a generalization of the six vertex current operator introduced in \cite{Odyssey}.
Expression (\ref{Bstr}) is certainly a symmetry operator as it maps degenerate subspaces onto itself.
But the symmetry operators which generalize the Chevalley operators found and studied in \cite{dfm}
are still elusive.\\

\noindent
{\large\bf Acknowledgment}\\
\noindent
The author is pleased to thank Prof. Barry McCoy for his interest in this work and for
many helpful comments.\\
\app
{Theta functions}
\label{Theta}
The definition of Jacobi Theta functions of nome $q$ (see (21.62) \cite{WhittW}) is
\begin{eqnarray}
\Hs(v)&=&2\sum_{n=1}^\infty(-1)^{n-1}q^{(n-{1/2})^2}\sin[(2n-1)\pi
v/(2K)]\\
\Theta(v)&=&1+2\sum_{n=1}^{\infty}(-1)^nq^{n^2}\cos(nv\pi/K)\nonumber\\
\label{thetadf}
\end{eqnarray}
where $K$  and $K'$ are the standard elliptic integrals of the first kind
and 
\begin{equation}
q=\exp{-\pi K'/ K}.
\end{equation}
These theta functions satisfy the quasi periodicity 
relations 
\begin{equation}
\Hs(v+2K)=-\Hs(v) \hspace{0.4 in}
\Hs(v+2iK')=-q^{-1}\exp(-\pi i v/K)\Hs(v)
\label{Hper}
\end{equation}
\begin{equation}
\Theta(v+2K)=~~\Theta(v) \hspace{0.4 in}
\Theta(v+2iK')=-q^{-1}\exp(-\pi i v/ K)\Theta(v).
\end{equation}
$\Theta(v)$ and $\Hs(v)$ are related by 
\begin{equation}
\Theta (v+iK')=iq^{-1/4}\exp{-\frac{\pi i v}{2K}}\Hs(v) 
\hspace{0.4 in}
\Hs(v+iK')=iq^{-1/4}\exp-{\frac{\pi i v}{2K}}\Theta(v).
\label{threl}
\end{equation}
They satisfy the addition theorem
\begin{equation}
\Hs(u)\Theta(v) - \Theta(u)\Hs(v)=2\Hs((u-v)/2)\Theta((u-v)/2)\Hs((u+v)/2+K)\Theta((u+v)/2+K)/(\Hs(K)\Theta(K))
\end{equation}

\app{The algebraic Bethe ansatz}
\label{app1}
We only list those definitions and identities which we make use of.
We follow the formalism of \cite{TakFadd79}.
The monodromy matrix is defined as
\begin{equation}
{\cal{T}} = {\cal{L}}_N \cdots {\cal{L}}_1
=
\left(\begin{array}{c c}
A & B \\
C & D \\
\end{array} \right)
\label{mono}
\end{equation}
where ${\cal{L}}_n$ is a $2 \times 2$ matrix 
in auxiliary space with entries which are
$2 \times 2$ matrices in spin space acting on 
the $n$ th spin in the spin chain and
$A,B,C,D$ are $2^{N}\times 2^{N}$ matrices in spin space.
\begin{equation}
{\cal{L}}_n = 
\left(\begin{array}{c c}
\alpha_n & \beta_n \\
\gamma_n & \delta_n\\
\end{array} \right)
\label{lT}
\end{equation}
\begin{equation}
\alpha_n = 
\left(\begin{array}{c c}
a & 0\\
0 & b  \\
\end{array} \right),
\hspace{0.1 in}
\beta_n = 
\left(\begin{array}{c c}
0&d\\
c&0  \\
\end{array} \right)
\hspace{0.1 in}
\gamma_n = 
\left(\begin{array}{c c}
0 &c\\
d & 0  \\
\end{array} \right),
\hspace{0.1 in}
\delta_n = 
\left(\begin{array}{c c}
b & 0\\
0 & a\\
\end{array} \right)
\label{ad}
\end{equation}
$a,b,c$ and $d$ are defined in (\ref{bw8}).
The R matrix is
\begin{equation}
{\cal{R}}^{\rho,\beta}_{\alpha,\kappa}(\lambda,\mu)=\frac{1}{2}(a_R+b_R)E \otimes E +
                              \frac{1}{2}(a_R-b_R){\sigma^3} \otimes {\sigma^3}+
                              c_R({\sigma^-}\otimes {\sigma^+}+{\sigma^+}\otimes{\sigma^-})+
                              d_R({\sigma^+}\otimes{\sigma^+}+{\sigma^-}\otimes{\sigma^-})
\label{R}
\end{equation}
where on the right hand side the first factor of the direct products carries the 
indices $\alpha,\rho$ and the second factor $\kappa,\beta$.
\begin{equation}
a_R = \Theta(2\eta)\Theta(\lambda-\mu)\Hs(\lambda-\mu+2\eta)
\end{equation}
\begin{equation}
b_R = \Hs(2\eta)\Theta(\lambda-\mu)\Theta(\lambda-\mu+2\eta)
\end{equation}
\begin{equation}
c_R = \Theta(2\eta)\Hs(\lambda-\mu)\Theta(\lambda-\mu+2\eta)
\end{equation}
\begin{equation}
d_R = \Hs(2\eta)\Hs(\lambda-\mu)\Hs(\lambda-\mu+2\eta)
\end{equation}
\begin{equation}
a = a_R(\lambda,\eta)~~~b = c_R(\lambda,\eta)~~~c = b_R(\lambda,\eta)~~~d = d_R(\lambda,\eta)
\end{equation}
and the local transition matrix (\ref{lT}),(\ref{ad}) is
\begin{equation}
{\cal{L}}^{\beta,\rho}_{\alpha,\kappa}(\lambda)={\cal{R}}^{\rho,\beta}_{\alpha,\kappa}(\lambda,\eta)
\end{equation}
In order to construct a generating vector suitable for the eight-vertex model
a gauge transformed {\mbox monodromy} matrix is defined as
\begin{equation}
{\cal{T}}_{k,l} = M^{-1}_{k}(\lambda) {\cal T(\lambda)} M_{l}(\lambda) =
\left(\begin{array}{c c}
A_{k,l} & B_{k,l} \\
C_{k,l} & D_{k,l} \\
\end{array} \right).
\label{tmono}
\end{equation}
where the  matrices $M_k$ introduced by Baxter \cite{bax732} are
\begin{equation}
M_k = 
\left(\begin{array}{c c}
x^{1}_{k} & y^{1}_{k}\\
x^{2}_{k} & y^{2}_{k}\\
\end{array} \right)
\hspace{0.4 in}
M^{-1}_k = \frac{1}{m(\lambda)}
\left(\begin{array}{c c}
~~y^2_{k}&  -y^1_{k}\\
-x^2_{k}&   ~~x^1_{k}\\
\end{array} \right)
\end{equation}
The columns of $M$ and rows of $M^{-1}$ , called intertwining vectors (\cite{DdVG}) are of central importance:
\begin{equation}
X_{k}(\lambda)=
\left(\begin{array}{c}
x^{1}_{k}(\lambda)\\
x^{2}_{k}(\lambda)\\
\end{array} \right)=
\left(\begin{array}{c}
\Hs(s+2k\eta-\lambda)\\
\Theta(s+2k\eta-\lambda)\\
\end{array} \right) \hspace{0.5 in} 
\label{Xk}
\end{equation}
\begin{equation}
Y_{k}(\lambda) = 
\left(\begin{array}{c}
y^{1}_{k}(\lambda)\\
y^{2}_{k}(\lambda)\\
\end{array} \right)=\frac{1}{g(\tau_{k})}
\left(\begin{array}{c}
\Hs(t+2k\eta+\lambda)\\
\Theta(t+2k\eta+\lambda)\\
\end{array} \right)
\label{Yk}
\end{equation}
\begin{equation}
 g(u) = \Hs(u)\Theta(u), \hspace{0,3 in} \tau_l=(s+t)/2+2l\eta-K
\end{equation}
\begin{equation}
\tilde{Y}_{k}(\lambda) = (~y^{2}_{k}(\lambda), -y^{1}_{k}(\lambda))
\hspace{0.3 in}
\tilde{X}_{k}(\lambda) = (-x^{2}_{k}(\lambda), ~x^{1}_{k}(\lambda))
\label{tyx}
\end{equation}
\begin{equation}
(\tilde{X}^{\mu}_{k}(\lambda)X_{k,\mu}(\lambda)) = 0
\hspace{0.4 in}
(\tilde{Y}^{\mu}_{k}(\lambda)Y_{k,\mu}(\lambda)) = 0
\label{XXYY}
\end{equation}
\begin{equation}
(\tilde{X}^{\mu}_{k}(\lambda)Y_{k,\mu}(\lambda)) = m(\lambda)
\hspace{0.3 in}
(\tilde{Y}^{\mu}_{k}(\lambda)X_{k,\mu}(\lambda)) = m(\lambda)
\label{XY}
\end{equation}
\begin{equation}
m(\lambda)=\frac{2g(\lambda+(t-s)/2)}{g(K)}
\end{equation}
The local vectors $X_k$ defined in (\ref{Xk}) generalize the local vacuum of the six  vertex model 
\begin{equation}
e^{+}=
\left(\begin{array}{c}
1\\
0\\
\end{array} \right)
\label{lV}
\end{equation}
to the eight vertex model
and build up the generating vectors (4.18) in \cite{TakFadd79} 
\begin{equation}
\Omega_{N}^{l} = \omega_1^l\otimes \cdots \otimes \omega_{N}^l
\label{Omega}
\end{equation}
\begin{equation}
\omega_{k}^{l} = X_{k+l}(\eta)
\label{omega}
\end{equation}
The elements of ${\cal{T}}_{k,l}$ are
\begin{equation}
A_{k,l}(\lambda) = \frac{1}{m(\lambda)}\tilde{Y}_{k}(\lambda){\cal{T}}(\lambda)X_{l}(\lambda)
\label{Aop}
\end{equation}
\begin{equation}
B_{k,l}(\lambda) = \frac{1}{m(\lambda)}\tilde{Y}_{k}(\lambda){\cal{T}}(\lambda)Y_{l}(\lambda)
\label{Bop}
\end{equation}
\begin{equation}
C_{k,l}(\lambda) =\frac{1}{m(\lambda)}\tilde{X}_{k}(\lambda){\cal{T}}(\lambda)X_{l}(\lambda)
\label{Cop}
\end{equation}
\begin{equation}
D_{k,l}(\lambda) =\frac{1}{m(\lambda)}\tilde{X}_{k}(\lambda){\cal{T}}(\lambda)Y_{l}(\lambda)
\label{Dop}
\end{equation}
The gauge transformed local transition matrix is
\begin{equation}
{\cal{L}}^{l} = M^{-1}_{l}(\lambda) {\cal L(\lambda)} M_{l-1}(\lambda) =
\left(\begin{array}{c c}
\alpha^{l} & \beta^{l} \\
\gamma^{l} & \delta^{l} \\
\end{array} \right).
\label{gL}
\end{equation}
\begin{equation}
\alpha^l(\lambda) = \frac{1}{m(\lambda)}\tilde{Y}^{\mu}_{l}(\lambda){\cal{L}}^{\beta,\nu}_{\alpha,\mu}(\lambda)X_{l-1,\nu}(\lambda)
=\frac{1}{m(\lambda)}\tilde{Y}^{\mu}_{l}(\lambda){\cal{R}}(\lambda,\eta)^{\nu,\beta}_{\alpha,\mu}X_{l-1,\nu}(\lambda)
\label{alpha}
\end{equation}
\begin{equation}
\beta^l(\lambda) = \frac{1}{m(\lambda)}\tilde{Y}^{\mu}_{l}(\lambda){\cal{L}}^{\beta,\nu}_{\alpha,\mu}(\lambda)Y_{l-1,\nu}(\lambda)
=\frac{1}{m(\lambda)}\tilde{Y}^{\mu}_{l}(\lambda){\cal{R}}(\lambda,\eta)^{\nu,\beta}_{\alpha,\mu}Y_{l-1,\nu}(\lambda)
\label{beta}
\end{equation}
\begin{equation}
\gamma^l(\lambda) = \frac{1}{m(\lambda)}\tilde{X}^{\mu}_{l}(\lambda){\cal{L}}^{\beta,\nu}_{\alpha,\mu}(\lambda)X_{l-1,\nu}(\lambda)
=\frac{1}{m(\lambda)}\tilde{X}^{\mu}_{l}(\lambda){\cal{R}}(\lambda,\eta)^{\nu,\beta}_{\alpha,\mu}X_{l-1,\nu}(\lambda)
\label{gamma}
\end{equation}
\begin{equation}
\delta^l(\lambda) = \frac{1}{m(\lambda)}\tilde{X}^{\mu}_{l}(\lambda){\cal{L}}^{\beta,\nu}_{\alpha,\mu}(\lambda)Y_{l-1,\nu}(\lambda)
=\frac{1}{m(\lambda)}\tilde{X}^{\mu}_{l}(\lambda){\cal{R}}(\lambda,\eta)^{\nu,\beta}_{\alpha,\mu}Y_{l-1,\nu}(\lambda)
\label{delta}
\end{equation}
{\bf Action of $\alpha^l,\beta^l,\gamma^l,\delta^l$}\\
To compute the left hand side of equ. (\ref{singlecol}) we need the relations
\begin{equation}
\delta^l(\lambda)X_{l}(\eta) = h(\lambda-\eta)X_{l+1}(\eta)
\label{deltaX}
\end{equation}
\begin{equation}
\beta^l(\lambda)X_{l}(\eta) = h(2\eta)\frac{g(\tau_l+\lambda-\eta)}{g(\tau_l)}Y_{l-1}(\eta)
\label{betaX}
\end{equation}
\begin{equation}
\delta^l(\lambda)Y_{l-2}(\eta) = h(\lambda+\eta)\frac{(\tilde{X}^{\mu}_{l}(\lambda)Y_{l-2,\mu}(\lambda))}{m(\lambda)}Y_{l-1}(\eta)
\label{deltaY}
\end{equation}
They follow from  (\ref{ITV1})-(\ref{ITV12}). Equ. (\ref{deltaX}) is identical with equ. (4.16) of \cite{TakFadd79}.\\
{\bf Intertwining vectors}\\
We make extensive use of the following powerful relations describing the action of the R-matrix on intertwining
vectors \cite{bax732}, \cite{TakFadd79}.
\begin{equation}
{\cal{R}}(\la,\mu)(X_{l}(\la) \otimes X_{l+1}(\mu)) = 
h(\la- \mu + 2\eta)X_{l}(\mu) \otimes X_{l+1}(\la)
 \label{ITV1}
\end{equation}
\begin{equation}
{\cal{R}}(\la,\mu)(Y_{l+1}(\la) \otimes Y_{l}(\mu)) = 
h(\la- \mu + 2\eta)Y_{l+1}(\mu) \otimes Y_{l}(\la)
\label{ITV2}
\end{equation}
\begin{equation}
{\cal{R}}(\la,\mu)(Y_{k}(\la) \otimes X_{l}(\mu)) = 
f^{YX}_{1,k,l}(\la- \mu)Y_{k}(\mu) \otimes X_{l}(\la)+
f^{YX}_{2,k,l}(\la- \mu)X_{l+1}(\mu) \otimes Y_{k+1}(\la)
\label{ITV3}
\end{equation}
\begin{equation}
{\cal{R}}(\la,\mu)(X_{k}(\la) \otimes Y_{l}(\mu)) = 
f^{XY}_{1,k,l}(\mu- \la)X_{k}(\mu) \otimes Y_{l}(\la)+
f^{XY}_{2,k,l}(\la- \mu)Y_{l-1}(\mu) \otimes X_{k-1}(\la)
\label{ITV4}
\end{equation}
\begin{equation}
(\tilde{Y}_{l}(\mu) \otimes \tilde{Y}_{l+1}(\la)){\cal{R}}(\la,\mu) = 
h(\la- \mu + 2\eta)\tilde{Y}_{l}(\la) \otimes \tilde{Y}_{l+1}(\mu)
\label{ITV5}
\end{equation}
\begin{equation}
(\tilde{X}_{l+1}(\mu) \otimes \tilde{X}_{l}(\la)){\cal{R}}(\la,\mu) = 
h(\la- \mu + 2\eta)\tilde{X}_{l+1}(\la) \otimes \tilde{X}_{l}(\mu)
\label{ITV6}
\end{equation}
\begin{equation}
(\tilde{X}_{k}(\mu) \otimes \tilde{Y}_{l}(\la)){\cal{R}}(\la,\mu) = 
f^{YX}_{1,l,k}(\lambda-\mu)\tilde{X}_{k}(\la) \otimes \tilde{Y}_{l}(\mu)
+f^{YX}_{2,l,k}(\lambda-\mu)\tilde{Y}_{l+1}(\la) \otimes \tilde{X}_{k+1}(\mu)
\label{ITV7}
\end{equation}
\begin{equation}
(\tilde{Y}_{k}(\mu) \otimes \tilde{X}_{l}(\la)){\cal{R}}(\la,\mu) = 
f^{XY}_{1,l,k}(\mu-\la)\tilde{Y}_{k}(\la) \otimes \tilde{X}_{l}(\mu)
+f^{XY}_{2,l,k}(\lambda-\mu)\tilde{X}_{l-1}(\la) \otimes \tilde{Y}_{k-1}(\mu)
\label{ITV8}
\end{equation}

with
\begin{equation}
f^{YX}_{1,k,l}(\la-\mu) = 
\frac{h(2\eta)g(\tau_{(k+l+1)/2}+\lambda-\mu)}
{g(\tau_{(k+l+1)/2})}
\label{ITV9}
\end{equation}
\begin{equation}
f^{YX}_{2,k,l}(\la- \mu) = 
\frac{h(\lambda-\mu)g(\tau_{(k+l-1)/2})g(\tau_{k+1})}
{g(\tau_{(k+l+1)/2})g(\tau_{k})} 
\label{ITV10}
\end{equation}
\begin{equation}
f^{XY}_{1,k,l}(\mu- \la) =
\frac{h(2\eta)g(\tau_{(k+l-1)/2}+\mu - \lambda)}
{g(\tau_{(k+l-1)/2})}
\label{ITV11}
\end{equation}
\begin{equation}
f^{XY}_{2,k,l}(\la- \mu)=
\frac{h(\lambda-\mu)g(\tau_{(k+l+1)/2})g(\tau_{l-1})}
{g(\tau_{(k+l-1)/2})g(\tau_{l})} 
\label{ITV12}
\end{equation}
{\bf Permutation relations for $A_{k,l},\cdots ,D_{k,l}$ }\\
\label{Perm}
${\cal{T}}$and ${\cal{R}}$ defined in (\ref{mono}) and (\ref{R}) satisfy the RTT equation
\begin{equation}
{\cal{R}}(\lambda,\mu)({\cal{T}}(\lambda)\otimes{\cal{T}}(\mu)) =
({\cal{T}}(\mu)\otimes{\cal{T}}(\lambda)){\cal{R}}(\lambda,\mu)
\label{YB}
\end{equation}
which combined with  (\ref{ITV1})-(\ref{ITV8}) leads to the following fundamental permutation relations:
\begin{equation}
B_{k,l+1}(\lambda)B_{k+1,l}(\mu)=B_{k,l+1}(\mu)B_{k+1,l}(\lambda)
\label{Bs}
\end{equation}
\begin{equation}
A_{k,l}(\lambda)B_{k+1,l-1}(\mu)=\alpha(\lambda,\mu)B_{k,l-2}(\mu)A_{k+1,l-1}(\lambda)
-\beta_{l-1}(\lambda,\mu)B_{k,l-2}(\lambda)A_{k+1,l-1}(\mu)
\label{C2}
\end{equation}
\begin{equation}
D_{k,l}(\lambda)B_{k+1,l-1}(\mu)=\alpha(\mu,\lambda)B_{k+2,l}(\mu)D_{k+1,l-1}(\lambda)
+\beta_{k+1}(\lambda,\mu)B_{k+2,l}(\lambda)D_{k+1,l-1}(\mu)
\label{C3}
\end{equation}
where
\begin{equation}
\alpha(\lambda,\mu)=\frac{h(\lambda-\mu-2\eta)}{h(\lambda-\mu)},
~~~{\rm and}~~~
\beta_k(\lambda,\mu)=\frac{h(2\eta)h(\tau_k+\mu-\lambda)}{h(\tau_k)h(\mu-\lambda)}
\label{calpha}
\end{equation}
and where
\begin{equation}
h(u)=\Theta(0)\Theta(u)\Hs(u)
\end{equation}
As follows from (\ref{ITV1})-(\ref{ITV12}) these relations continue to be formally valid for a chain of length zero.
We will need equ. (\ref{C2}) in this limit:
\begin{eqnarray}
\label{C20}
&&(\tilde{Y}_{k}(\lambda)X_{l}(\lambda))(\tilde{Y}_{k+1}(\mu)Y_{l-1}(\mu))=  \\
&&\alpha(\lambda,\mu)(\tilde{Y}_{k}(\mu)Y_{l-2}(\mu))(\tilde{Y}_{k+1}(\lambda)X_{l-1}(\lambda))
-\beta_{l-1}(\lambda,\mu)(\tilde{Y}_{k}(\lambda)Y_{l-2}(\lambda))(\tilde{Y}_{k}(\mu)X_{l-1}(\mu)) \nonumber
\end{eqnarray}

\app{The string operator.}
\label{ZXYP}
The string operator (\ref{Bstr}) found in \cite{ellc} is written in terms of the quantities
\begin{equation}
{\hat Z}_1(\lambda_c)=\frac{{\hat X}(\lambda_c)}{{\hat Y}(\lambda_c)}
\label{X}
\end{equation}
with
\begin{eqnarray}
&&{\hat X}(\lambda_c)=-2\sum_{k=0}^{L_s-1}k
\frac{\omega^{-2(k+1)}\rho_{k+1}}{P_kP_{k+1}}\\
&&{\hat Y}(\lambda_c)=
\sum_{k=0}^{L_s-1}\frac{\omega^{-2(k+1)}\rho_{k+1}}{P_kP_{k+1}}
\label{Y}
\end{eqnarray}
and
\begin{equation}
\hat{Z}_{j}(\lambda_c)=\hat{Z}_1(\lambda_c-(j-1)2\eta)
\end{equation}
where 
$\omega=e^{2\pi i m/L}$ is a L th root of unity. 
\begin{equation}
\rho_k = h^{N}(\lambda_c-(2k-1)\eta)
\hspace{0.4 in}
P_k = \prod_{m=1}^{n_r}h(\lambda_c-\lambda^{r}_{m}-2k\eta).
\label{P}
\end{equation}

$\lambda^{r}_{m}, m=1,\cdots,n_r$ denote regular Bethe-roots, $\lambda_c$ denotes the string center.

\end{document}